\documentstyle[mprocl]{article}

\def\be{\begin{equation}}
\def\ee{\end{equation}}
\def\bea{\begin{eqnarray}}
\def\eea{\end{eqnarray}}

\begin{document}

\title{RELATIVISTIC STUDIES OF CLOSE NEUTRON STAR BINARIES}

\author{ G.J. MATHEWS,  P. MARRONETTI}

\address{University of Notre Dame, Department of Physics, 
Notre Dame, IN 46556, USA}

\author{ J.R. WILSON}

\address{Lawrence Livermore National Laboratory, Livermore, CA 94550, and \\
University of Notre Dame, Department of Physics, Notre Dame, IN 46556, USA}


\maketitle\abstracts{
We discuss (3+1) dimensional general
relativistic hydrodynamic simulations of close neutron star
binary systems.  The relativistic
field equations are solved at each time slice with a spatial 3-metric
chosen to  be conformally flat.  Against this solution
the hydrodynamic variables and gravitational radiation are allowed
to respond.  We have studied  four physical processes which occur
as the stars approach merger.  These include:
1) the relaxation to a hydrodynamic state of almost no spin; 2)
relativistically driven compression, heating, and neutrino
emission; 3) collapse to  two black holes; and 4) orbit
inspiral occurring at a lower frequency than previously expected.
We give a brief account of the physical origin of these effects
and an explanation of why they do not appear in models based
upon, 1PN hydrodynamics, a weak field multipole expansion,
a tidal analysis, or a rigidly corotating velocity field.
The implication of these results for gravity wave detectors is
also discussed.}

The physical processes occurring during the last orbits of
a neutron star binary are currently a subject of intense 
interest.\cite{wm95}$^-$\cite{baumgarte}
In part, the recent surge in interest stems from
relativistic numerical hydrodynamic simulations
in which it has been noted \cite{wm95,wmm96,mw97}  that as the
stars approach their final orbits they 
experience compression. Indeed,
for an appropriate equation of state,
numerical simulations \cite{mw97} indicate that binary neutron 
stars heat significantly before
individually collapsing  toward black holes
many seconds prior to merger.  The orbit frequency is also
significantly lower than that of Newtonian or post-Newtonian
point sources, and the inner most stable orbit occurs at a 
larger separation distance.\cite{wmm96}  All of these effects could
have a significant impact on the anticipated
gravity wave signal from merging neutron stars.  They could
also provide an energy source for cosmological gamma-ray 
bursts.\cite{mw97}
 
However, a number of recent papers
\cite{lai}$^-$\cite{baumgarte} have
not observed this effect in \break
Newtonian,\cite{lai} 
1PN,\cite{shabata,wiseman} weak field multipole
 expansions,\cite{brady,flanagan,thorne} or in binaries
in which rigid corotation has been imposed.\cite{baumgarte}
Moreover, this flurry of activity has caused some confusion as
to the physics responsible for the effects observed in the numerical
calculations.  Here, we present a brief 
derivation of the physics which drives
the compression and discuss how such terms are missed
in the various approximation schemes.
We describe simulations  which demonstrate that  
the compression forces do not appear in 
simple linear motion or rigid corotation.
We also summarize the implications of these results on the 
gravity wave signal of close neutron star binaries.

The basic physical
processes which induce compression can be traced to completely general terms
in the hydrodynamic equations of motion.\cite{mw97}
We begin with the usual ADM (3+1) metric \cite{adm62,york79}
in which there is a slicing of the spacetime into a one-parameter
family of  three-dimensional hypersurfaces $\gamma_{i j}$
separated by differential displacements
in a time-like coordinate $\alpha$,
\begin{equation}
ds^2 = -(\alpha^2 - \beta_i\beta^i) dt^2 +
2 \beta_i dx^i dt + \gamma_{ij}dx^i dx^j~~.
\label{metric}
\end{equation}
The conformally flat metric condition  ({\it CFC}) expresses the
three metric of Eq. (\ref{metric}) as 
$\gamma_{i j} =  \phi^4 \delta_{ij}$.
It is common practice
to impose this condition when solving the initial value problem
in numerical relativity (which is in essence what we do).
One question, however, is the amount of hidden radiation \cite{abrahams}
contained in the CFC solution.
We have estimated this 
by decomposing the extrinsic curvature
into longitudinal $K^{i j}_L$ and
transverse $K^{i j}_T$ components as proposed by York.\cite{york73}
By this order-of-magnitude estimate, we find that the
"hidden" gravitational radiation
energy density is a small fraction of the total gravitational
mass energy of the system, $
\int  K^{i j}_T K_{Tij} {dV\over 8 \pi} \approx 2 \times 10^{-5} ~
{\rm M_G}$. Similarly, the multipole estimate of the 
power loss in gravitational radiation
is a small fraction of the energy in orbital motion
 $\dot J/\omega J\sim 10^{-4}$.
Hence, the CFC is probably a good approximation to the initial data
for the binaries we study.

The vanishing of the
spatial components of the divergence of the
energy momentum tensor
$(T_\mu^{~i})_{;\mu} = 0$
leads to an evolution equation for the 
covariant four momentum,
\begin{eqnarray}
{\dot S_i}& + &  S_i{\dot \gamma \over\gamma}
-{1\over\gamma}{\partial\over\partial x^j}(S_iV^j\gamma)
 + {\alpha \partial P\over \partial x^i}
- S_j {\partial \beta^j \over \partial x^i} \nonumber \\
&+ & (\rho(1 + \epsilon) + P)\biggl(W^2{\partial \alpha \over \partial x^i}
+ \alpha {U_j U_k \over 2 } {\partial \gamma^{j k}
\over \partial x^i}\biggr) = 0~~.
\label{hydromom}
\end{eqnarray}
In an orbiting system it is convenient to allow $\beta^j$ to follow the
the orbital motion of the stars. In which case, the
 term $S_j (\partial \beta^j /\partial x^i)$ contains the 
centrifugal force (plus some small frame drag).  The term containing
$\partial \alpha/\partial x^i$ is the analog of the
Newtonian gravitational force.
 
The term  with $(U_j U_k /2 ) \partial \gamma^{j k}
/\partial x^i$ is the compression driving force.
It does not have a Newtonian analog.
This term vanishes for a star at rest with respect
to the observer or in the flat-space limit.  However,
for a star with fluid motion  in curved space, it
describes an additional force represented as a product of
velocities times the gradient of the three metric.
For simple linear motion the effects of
this term should cancel to leave the
stellar structure unchanged.  Similarly, this term appears
to cancel \cite{baumgarte} for fluid motion
in which the four velocity can be taken as proportional
to a simple  Killing vector.  However,
for more general states of motion (e.g.~noncorotating stars,
differential rotation, meridional
circulation, turbulent flow, etc.)  the effects of this force must
be evaluated numerically.  Indeed, the sign of this force is such that
a lower energy configuration for a binary star than that of rigid
corotation is obtained by allowing the fluid to respond
to this force term.  We find \cite{wm98} that the numerical
relaxation  of binary stars from corotation (or any other
initial spin configuration) produces
a state of almost no net spin in which the central density
and gravitational binding energy increase.

We have performed numerous numerical tests which substantiate
that this term does
indeed vanish when the hydrodynamics is artificially
constrained to uniform translation, stationary stars in a tidal
field, or approximate rigid corotation.  Indeed,
 the constrained stars remain near the central density
of an isolated star, while the binary stars show an increase in
central density which
grows as $\sim(v/c)^4$.
We have also  analyzed \cite{wm98} the nature
and formation of this state in more detail by
imposing an initial  rigid angular velocity
in the frame of the orbiting stars
in the
range  $-900 < \omega_S < 900$ rad sec$^{-1}$, corresponding
to $-0.03 < J_S/m_0^2 < 0.12$.  
In each case, the stars relax to
a state of almost no net spin within about
three sound crossing times ($\sim 0.6$ msec).

We  have also  used a multipole expansion \cite{wmm96} to extract
the gravity wave signal.
A striking feature of these simulations is that
the frequency varies slowly as the orbit decays (less chirp).
This imples a higher signal to noise for gravity wave detectors
at low frequencies ($\sim 100$ Hz).
At least part of the differences with PN estimates \cite{kidder}
can be attributed to the
effects of finite stellar size.\cite{shabata}   However, we attribute  most
of this difference to time dilation      
and length contraction in the strong field of the binary.\cite{wmm96}
As the stars collapse, an abrupt change in the gravity wave frequency might 
also be detected.  
Furthermore, the orbit instability can occur when
the specific orbital angular momentum is
in excess of unity.\cite{wmm96} Hence,
a very large amount of gravity wave emission
may accompany the final merger to a single Kerr black hole.

\section*{Acknowledgments}
 
Work at University of Notre Dame
supported in part by DOE Nuclear Theory DE-FG02-95ER40934,
NSF PHY-97-22086, and by NASA CGRO NAG5-3818.
Work performed in part under the auspices
of the U.~S.~Department of Energy
by the Lawrence Livermore National Laboratory under contract
W-7405-ENG-48 and NSF grant PHY-9401636.

\end{document}